\documentclass[sigconf]{acmart}
\usepackage{xcolor}
\usepackage{enumitem}
\usepackage{cleveref}

\usepackage{listings}

\lstdefinestyle{pythonstyle}{
  language=Python,
  basicstyle=\ttfamily\footnotesize,
  keywordstyle=\color{blue},
  commentstyle=\color{gray},
  stringstyle=\color{orange!90!black},
  showstringspaces=false,
  tabsize=2,
  frame=single,
  breaklines=true,
  linewidth=0.47\textwidth,
  numbers=left,                    
  numbersep=5pt,                   
  stepnumber=1                     
}

\AtBeginDocument{%
  }

\setcopyright{none} 
\copyrightyear{2026}
\acmYear{2026}
\acmDOI{XXXXXXX.XXXXXXX}

\acmConference[MICRO 2026]{The 58th IEEE/ACM International Symposium on Microarchitecture}{October 31, 2026--November 4, 2026}{Athens, Greece}

\acmISBN{978-X-XXXX-XXXX-X/XX/XX}

\setcopyright{none}

\renewcommand\footnotetextcopyrightpermission[1]{}

\newcommand{\parhead}[1]{\vspace{0.2em}\noindent\textbf{#1}}
\newcommand{\ignore}[1]{}

\newcommand{\papername}{Synchronization-Free All-Reduce}
\newcommand{\shortname}{SiFAR}
\newcommand{\Shortname}{\shortname\ }
\newcommand{\fullname}{\papername\ (\shortname)}

\newcommand{\ar}{All-Reduce}
\newcommand{\AR}{\ar\ }

\newcommand{\us}{$\mu$s\ }

\newcommand\new[1]{{\color{black}#1}}
\newcommand\replace[2]{#2}
\begin{document}

\title[SiFAR: Synchronization-Free All-Reduce for Low-Latency LLM Inference]{SiFAR: Synchronization-Free All-Reduce \\ for Low-Latency LLM Inference}

\author{%
  \begin{tabular}{@{}cccccc@{}}
    \shortstack{
      {\large\strut Hritvik Taneja\textsuperscript{*}}\\[-0.25em]
      {\small\strut htaneja3@gatech.edu}
    }
    &
    \shortstack{
      {\large\strut Anish Saxena\textsuperscript{*}}\\[-0.25em]
      {\small\strut asaxena317@gatech.edu}
    }
    &
    \shortstack{
      {\large\strut Abhishek Revinipati\textsuperscript{*}}\\[-0.25em]
      {\small\strut arevinipati3@gatech.edu}
    }
    &
    \shortstack{
      {\large\strut Jae Hyung Ju\textsuperscript{*}}\\[-0.25em]
      {\small\strut jju35@gatech.edu}
    }
    &
    \shortstack{
      {\large\strut Neal C. Crago\textsuperscript{\textdagger}}\\[-0.25em]
      {\small\strut ncrago@nvidia.com}
    }
    &
    \shortstack{
      {\large\strut Moinuddin Qureshi\textsuperscript{*}}\\[-0.25em]
      {\small\strut moin@gatech.edu}
    }
  \end{tabular}
}

\renewcommand{\shortauthors}{Taneja et al.}






\begin{abstract}
The rise of reasoning models and agentic systems has made LLM token-generation latency a key bottleneck. 
Unlike chatbots, where latency gains saturate at human reading speed, these systems generate intermediate reasoning tokens that are not consumed by humans. As a result, per-token latency directly determines end-to-end response time.
To reduce latency, inference engines operate with minimal batching, making token generation bandwidth-bound. Tensor Parallelism addresses this bottleneck by sharding model weights across GPUs and loading them in parallel. However, scaling to more GPUs introduces \AR overheads that grow with GPU count. We observe that removing \AR improves token throughput by \replace{35\%}{43\%} for Llama-3.1-8B on 8 H200 GPUs.


To address this overhead, we propose \emph{\fullname}, which reduces synchronization overheads incurred by \AR during low-latency inference.
Existing oneshot and twoshot algorithms incur high synchronization overheads due to barriers before and after communication, accounting for 32–62\% of the latency for small payloads we observe in low-batch serving.
To tackle this, first, we find that the \emph{bottom barrier} in oneshot enforces a WAW dependency, and eliminate it by co-designing communication and model execution to enable dual buffering. 
However, oneshot scales poorly with GPU count. Twoshot performs better at higher TP degrees, but incurs an unavoidable bottom barrier.
To overcome this, we leverage in-switch reduction in modern switches, primarily used for reduce-scatter. We propose redundant pull, where each GPU reduces the entire all-reduce payload at the switch. This improves oneshot scalability while retaining its no bottom barrier advantage.
Finally, to reduce \emph{top barrier} overhead, we observe that each token generation step issues multiple \AR operations, keeping GPUs tightly synchronized after the first \ar.
We therefore propose \emph{speculative reduction}, which initiates data transfer without waiting for the top barrier and ensures correctness via a lightweight validation mechanism.
\Shortname reduces \AR latency by up to 52\% and improves end-to-end throughput by \replace{15.2\%}{18.6\%} for Llama-3.1-8B and \replace{12.2\%}{13.1\%} for Qwen3.5-397B-17B at TP=8.
\end{abstract}


\maketitle
\begingroup
\renewcommand{\thefootnote}{\fnsymbol{footnote}}
\footnotetext[1]{Georgia Institute of Technology, Atlanta, Georgia, USA}
\footnotetext[2]{NVIDIA, Santa Clara, California, USA}
\endgroup
\thispagestyle{plain}
\pagestyle{plain}
\section{Introduction}
The advent of agentic systems~\cite{muennighoff2025s1, li2025s, acharya2025agentic} and reasoning models~\cite{deepseekr1} has made low-latency LLM token generation critical. In human-facing LLM applications such as chatbots, each token is read by users, so reducing latency beyond human reading speed provides diminishing returns. In contrast, these systems generate intermediate reasoning tokens that are not consumed by humans, so per-token latency compounds across steps and directly impacts end-to-end response time.
Recent proposals for hardware accelerators tailored for reasoning~\cite{adiletta2026rpu} and industry platforms like NVIDIA Groq LPU~\cite{nvidia2026groq3lpx} further emphasize the importance of optimizing for low-latency inference. In this work, we focus on reducing token generation latency in GPU-based inference.

\parhead{The Low-Latency Challenge.}
The first step towards reducing token generation latency is to minimize batching, since larger batches delay individual token generation by (1) increasing the size of KV Cache and (2) activating more model parameters~(for MoE~\cite{deepseekv3}). With minimal batching, loading model weights and KV Cache from HBM becomes the primary bottleneck. Tensor Parallelism (TP) addresses this memory-bandwidth bottleneck by partitioning the model weights and KV Cache across multiple GPUs and loading them in parallel with aggregate bandwidth. However, scaling across GPUs to improve bandwidth introduces additional bottlenecks, including kernel launch overheads~\cite{blackwellpdl}, on-chip delays~\cite{jin2024uncovering}, HBM under-utilization~\cite{spector2025megakernel}, and communication overheads.

\parhead{All-Reduce: The Emerging Bottleneck.}
Recent advances in system software mitigate many non-communication bottlenecks. Megakernels~\cite{wu2025mirage,spector2025megakernel} eliminate kernel launch overheads and improve HBM utilization by fusing the forward pass, while Programmatic Dependent Launch~\cite{pdl_docs} enables prefetching across kernels. As these optimizations reduce compute-side overheads, \AR communication emerges as the dominant cost.

To quantify this trend, we measure time per output token (TPOT) for Llama~3.1~8B on H200 GPUs using a Megakernel implementation~\cite{spector2025megakernel}. As we scale from TP=1 to TP=8, TPOT reduces from \replace{2.8}{2.73}~ms to \replace{1.63}{1.43}~ms, and the fraction of time in \AR increases from 0\% to \replace{26\%}{30\%}~(corresponding to a \replace{35\%}{43\%} throughput gain). Thus, \AR becomes a growing contributor to latency and is poised to dominate as other bottlenecks diminish. Despite this, optimizing \AR for low-latency inference remains largely unexplored.

\begin{figure*}[htb!]
    \centering
    \includegraphics[width=\textwidth]{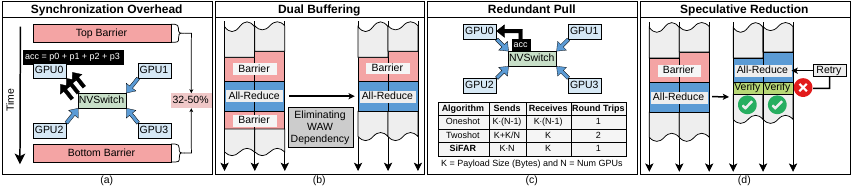}
    \caption{(a) Oneshot \AR with top and bottom barriers around data transfer; these barriers account for up to 50\% of latency for small payloads. (b) Dual buffering removes the bottom barrier by eliminating the write-after-write dependency. (c) Redundant pull uses in-switch reduction to reduce the full payload per GPU, extending its use beyond reduce-scatter in twoshot. (d) \Shortname reduces the top barrier overhead by using speculation and retrying on mis-speculation.}
    \label{fig:overview}
    \vspace{-0.7em}
\end{figure*}

\parhead{Limitations of Prior Work.}
Prior approaches to reducing \AR overhead, such as kernel fusion~\cite{chang2024flux,zheng2025tilelink,zheng2025tritondistributed} and microbatching~\cite{gond2025tokenweave,deepep2025,zhu2025nanoflow}, rely on overlapping communication with computation from large batches. However, these techniques are ineffective for low-batch inference, where there is insufficient compute to hide communication. The \textbf{goal} of this work is to directly optimize \AR to minimize its latency.

\parhead{All-Reduce Algorithms.}
The two most common \AR algorithms in LLM inference are \emph{oneshot}~\cite{vllm-custom-ar} and \emph{twoshot}~\cite{wang2025symmetricmemory}. Oneshot completes reduction in a single round trip, where each GPU pulls the entire buffer from all peers and reduces it locally. Twoshot decomposes reduction into a reduce-scatter followed by a broadcast: GPUs first reduce disjoint $K/N$ portions of the payload ($K$: payload size, $N$: number of GPUs), then broadcast the results so each GPU receives the full reduced output. Compared to oneshot, this reduces data transfer but requires two round trips.

\parhead{Key Findings.}
Both oneshot and twoshot require a synchronization barrier before and after the data transfer phase. We find that for small payloads typical of low-batch inference, these barriers account for 32-50\% of oneshot latency and 49-62\% of twoshot latency~(\Cref{fig:overview}~(a)). Our key insight is to minimize \AR overheads by reducing the cost of these expensive barriers. To that end, we propose \emph{\fullname}\footnote{\emph{Sifar} means zero in Urdu, reflecting our goal of making \AR overheads ``zero''.}.

\parhead{Removing Bottom Barrier via Dual Buffering.}
The bottom barrier in oneshot ensures all peers have finished reading the payload buffer before the owning GPU overwrites it with new data, enforcing a write-after-write dependency. We eliminate this barrier using dual buffering. The inference engine alternates between two payload buffers across successive \AR operations, ensuring that the payload is never overwritten before all peers finish reading. While dual buffering is well known, it is not widely used in current LLM serving systems because communication libraries like NCCL~\cite{nccl} are decoupled from model execution and cannot make assumptions about buffer reuse patterns. \Shortname overcomes this by co-designing communication with model execution, enabling dual buffering in systems like vLLM~\cite{vllm} and Megakernels~\cite{spector2025megakernel} without changes to their memory management~(\Cref{fig:overview}~(b)).

However, removing the bottom barrier is insufficient. Oneshot scales poorly with GPU count, making it slower than twoshot beyond TP=2. At TP=8 with an 8~KB payload, oneshot takes 4.01~\us for data transfer compared to 1.96~\us for twoshot. While twoshot has lower data transfer latency, it requires a bottom barrier that cannot be eliminated because it enforces a read-after-write dependency after the broadcast phase of twoshot. Ideally, we want oneshot latency to be comparable to twoshot so that dual buffering is effective.

\parhead{Reducing Data Transfer via In-Switch Computation.}
To achieve this, we leverage in-switch computation available in modern NVIDIA switches via the \texttt{multimem.ld\_reduce} instruction~\cite{ptxisa}. When a GPU issues \texttt{ld\_reduce}, the switch pulls data from all peers, performs the reduction, and returns the result to the requesting GPU. In existing systems, this primitive is primarily used in the reduce-scatter phase of twoshot, where each GPU reduces a disjoint $K/N$ portion of the payload at the switch.
However, we find that \texttt{ld\_reduce} can reduce oneshot latency beyond its conventional use in the reduce-scatter phase of twoshot. Reduce-scatter latency remains flat up to 64~KB and grows only slightly up to 256~KB. At TP=8, a 256~KB reduce-scatter means each GPU reduces 32~KB at the switch, which is comparable to the 4-32~KB payloads in low-batch inference.
Our insight is that if the switch can reduce 32~KB per GPU with low latency, each GPU can issue \texttt{ld\_reduce} over the entire 4–32~KB payload instead of a disjoint partition. 
We call this \emph{redundant pull}: each GPU reduces the full $K$-byte payload at the switch and receives only $K$ bytes instead of $K \times (N-1)$ in oneshot, ensuring dual buffering remains effective~(\Cref{fig:overview}(c)).



\parhead{Reducing Top Barrier Overhead via Speculative Reduction.}
The top barrier ensures all GPUs have produced valid partials before communication begins, enforcing a RAW dependency that cannot be simply eliminated. 
However, we observe that LLM decoding issues multiple \AR operations per token. As a result, GPUs remain tightly synchronized after the first \ar, as they execute identical operations across successive \ar s.
Thus, the overhead of the top barrier is primarily due to flag exchange, not divergence between GPUs. This leads to our third optimization: \emph{speculate and verify}. When a GPU enters an \ar, it speculatively begins fetching data from peers, assuming their data is ready, thereby avoiding explicit flag exchange. To ensure correctness, each GPU writes a flag to a validation buffer that is reduced alongside the payload. If the result equals $N \times \text{flag}$ after reduction, all GPUs had valid data and speculation succeeded. Otherwise, \Shortname reruns the \ar. In practice, speculative reduction substantially reduces top-barrier overhead~(\Cref{fig:overview}(d)).

\parhead{Results.}
We integrate \Shortname into Megakernels~\cite{spector2025megakernel}, a state-of-the-art for low-latency LLM inference, and evaluate it on H200s. \Shortname reduces \AR latency by up to 52\% for payloads up to 32~KB and improves throughput by \replace{15.2\%}{18.6\%} for Llama-3.1-8B and \replace{12.2\%}{13.1\%} for Qwen3.5-397B-17B at TP=8. Our contributions are as follows:
\begin{enumerate}
[leftmargin=*, labelwidth=0.3cm, labelsep=0.2cm, itemsep=0.0cm, topsep=0.05cm]
    \item We identify that synchronization barriers dominate \AR latency in low-latency tensor-parallel LLM inference, accounting for up to 62\% of total \AR time.
    \item We propose \fullname, which eliminates the bottom barrier via dual buffering, reduces data transfer via redundant pull, and minimizes the top barrier via speculative reduction.
    \item We demonstrate \Shortname reduces \AR latency by up to 52\% and improves end-to-end throughput by up to \replace{15.2\%}{18.6\%}.
\end{enumerate}

\clearpage
\section{Background and Motivation}

\subsection{LLM Inference: System Design and Metrics}
LLM inference consists of two phases: \textit{prefill} and \textit{decode}. During the prefill phase, the model processes all input prompt tokens in parallel, making it compute-bound. In contrast, the decode phase loads all the model weights from memory to produce the next token, making it memory bandwidth-bound. To amortize the cost of repeatedly loading model weights, most LLM inference engines batch requests from multiple users together during inference, trading off latency for higher aggregate throughput.

\parhead{Performance Metrics.}
The performance of LLM inference systems is measured using two latency metrics: \emph{Time-to-First-Token (TTFT)} and \emph{Time-per-Output-Token (TPOT)}. TTFT measures the time taken to process the prompt and produce the first output token, while TPOT measures how quickly subsequent tokens are generated. Both metrics are critical for user experience: TTFT determines the initial responsiveness, and TPOT governs the perceived generation speed. In this work, we focus on optimizing TPOT, as it dominates end-to-end latency in many use cases of LLM-based applications.

\subsection{Low Latency LLM Serving}
Traditional LLM-based applications, such as chatbots, are designed to be human-facing, where each generated token is read by a human. In such settings, reducing token generation latency (TPOT) beyond a certain point yields diminishing returns, since humans can only read 4-7 tokens/second~\cite{brysbaert2019many}. However, newer reasoning models~\cite{deepseekr1} and agentic systems~\cite{muennighoff2025s1, li2025s, acharya2025agentic} operate under a different paradigm. Here, tokens are either consumed by other LLMs or used internally as intermediate reasoning steps rather than being read by humans. As a result, per-token latency compounds across steps, making TPOT the dominant contributor to end-to-end response time. Industry platforms such as the NVIDIA Groq LPU~\cite{nvidia2026groq3lpx}, which stores model weights in on-chip SRAM to improve memory bandwidth, as well as recent hardware proposals for reasoning~\cite{adiletta2026rpu}, further highlight the growing importance of fast token generation. In this work, we focus on reducing TPOT in GPU-based inference.

\subsection{Techniques for Low-Latency Serving}
LLM inference systems face a fundamental tradeoff between throughput and interactivity. On one end, high-throughput systems maximize aggregate tokens generated per second across all users by batching hundreds of requests together. On the other end, low-latency systems minimize TPOT by reducing batch sizes, prioritizing responsiveness over throughput~\cite{inferencex}.

Batching requests from multiple users amortizes the cost of loading model weights across tokens. However, batching also increases token generation latency: larger batches require loading more KV cache during attention and activating more parameters in MoE models. Prior work~\cite{davies2025efficient} shows that TPOT increases substantially as batch size grows. We observe a similar trend when running Llama-3.1-8B and Qwen3.5-397B-17B at TP=8 on H200 GPUs using vLLM 0.18.1, where TPOT increases by more than 3.5$\times$ when scaling from BS=1 to BS=256. Thus, minimizing batch size is essential for reducing TPOT in latency-sensitive settings.

\begin{figure}[htb!]
\centering
\includegraphics[width=0.98\linewidth]{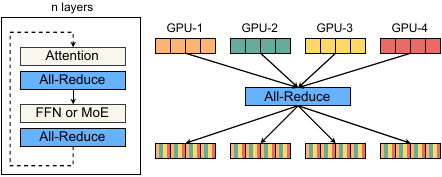}
\caption{Overview of TP and \ar: TP adds two \ar s to every layer of an LLM model. During \ar, each GPU pulls partial results from all peers and reduces them locally to produce the complete result.}
\label{fig:tp-and-ar}
\end{figure}

\parhead{Tensor Parallelism (TP) to Reduce TPOT.}
With minimal batching, autoregressive decoding becomes memory bandwidth-bound, since each token generation step repeatedly streams large model weights from GPU memory. TP~\cite{shoeybi2019megatron} is traditionally used to fit models that exceed the memory capacity of a single GPU by sharding weight matrices across devices. However, TP also increases aggregate memory bandwidth, as each GPU loads only a fraction of the weights. Thus, after reducing batching, TP can further improve TPOT for bandwidth-bound LLM inference, \new{even when the model fits on one GPU}. \Cref{fig:tp-and-ar} illustrates how TP operates in LLMs. TP shards weight matrices across GPUs, requiring each layer to perform two \AR operations: one after attention and one after the feed-forward network. In each \AR kernel all GPUs exchange partial results and reduce them to produce the final output.

Other parallelism strategies are less effective for reducing TPOT. Pipeline parallelism~\cite{huang2019gpipe} does not increase memory bandwidth, as each stage still loads full model weights sequentially. Expert parallelism~\cite{fedus2022switch} introduces load imbalance across experts, leading to uneven bandwidth utilization. As a result, neither approach is optimal for reducing TPOT. Thus, in this work, we focus on TP.

\subsection{Communication Bottlenecks in Tensor Parallelism (TP)}
Improving TPOT by scaling to more GPUs via TP introduces several overheads, including communication overheads from \ar, kernel launch overheads, low HBM utilization, and GPU on-chip latency~\cite{spector2025megakernel}. These overheads prevent ideal linear scaling of TPOT with increasing GPU count, which would otherwise be expected for bandwidth-bound, non-communication kernels. Recent system optimizations mitigate many of these non-communication bottlenecks. For example, Megakernels~\cite{wu2025mirage,spector2025megakernel} fuse the forward pass into a single kernel to eliminate launch overheads and improve HBM utilization, while Programmatic Dependent Launch (PDL)~\cite{pdl_docs} enables weights prefetching across kernel boundaries. Thus, as non-communication overheads are mitigated, \AR becomes the dominant bottleneck in low-latency inference.

\begin{figure}[htb!]
\centering
\includegraphics[width=\linewidth]{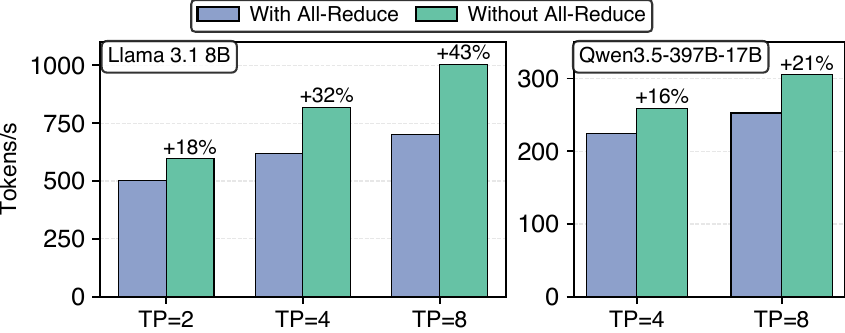}
\caption{\new{Impact of \AR on TPOT across different TP degrees for Llama-3.1-8B and Qwen3.5-397B-17B. At higher TP, removing \AR leads to larger gains, indicating that \AR becomes an increasingly dominant bottleneck.}}
\label{fig:allreduce_breakdown}
\vspace{-1em}
\end{figure}

To quantify the impact of \AR on TPOT, we use a Megakernel~\cite{spector2025megakernel} implementation, which fuses the forward pass into a single CUDA kernel, to evaluate Llama-3.1-8B~\cite{llama3}, and \new{Qwen3.5-397B-17B~\cite{qwen3.5}}. \Cref{fig:allreduce_breakdown} shows throughput (tokens/s) across different TP degrees, comparing execution with and without \ar~(i.e., \new{we feed the input of the \AR directly to the next operation instead}). For Llama-3.1-8B, removing \AR improves performance by \replace{16\%}{18\%} at TP=2 and \replace{35\%}{43\%} at TP=8. \new{For Qwen3.5-397B-17B, removing \AR improves performance by 16\% at TP=4 and 21\% at TP=8. Thus, using TP to achieve lower TPOT makes \AR a significant bottleneck and will become increasingly dominant as other overheads are reduced.} 



\subsection{Limitations of Prior Work}
\label{sec:prior-works}

\parhead{Microbatching.}
Prior works such as \emph{TokenWeave}~\cite{gond2025tokenweave}, \emph{DeepEP}~\cite{deepep2025}, and \emph{NanoFlow}~\cite{zhu2025nanoflow} use microbatching to overlap computation and communication by dividing large batches into smaller microbatches and pipelining their execution. This approach is effective at overcoming communication latencies when the workload is compute-bound, which occurs during prefill or decode with large batch sizes. However, in low-latency inference, batch sizes are intentionally kept small to minimize latency. As a result, microbatching provides little benefit since there is insufficient computation to overlap with communication. Moreover, microbatching primarily improves throughput at the expense of single token latency, which is the key objective in our setting.

\begin{figure}[htb!]
\centering
\includegraphics[width=\linewidth]{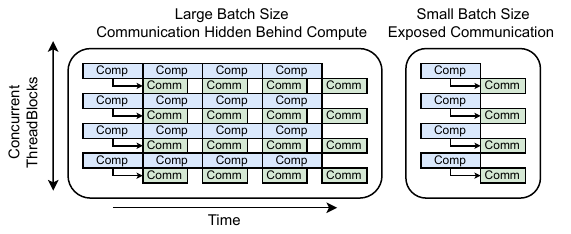}
\caption{Fine-grained compute-communication overlap. While effective at hiding communication overhead for large batch sizes, this technique provides minimal overlap in low-latency inference due to insufficient computation.}
\label{fig:kernel-fusion}
\end{figure}

\parhead{Fine-Grained Compute-Communication Overlap.}
Kernel fusion techniques~\cite{chang2024flux,zheng2025tilelink,zheng2025tritondistributed} fuse the \AR kernel to the preceding GEMM~(matrix multiply) kernel and overlap compute and communication by sending partial results tile-by-tile. While effective in hiding \AR overheads for compute-bound GEMMs with large batch sizes~(see \Cref{fig:kernel-fusion}~(left)), fusion provides little benefit in low-latency inference~(see \Cref{fig:kernel-fusion}~(right)). With small batches, each GEMM only issues a small number of threadblocks, which can execute concurrently on the GPU, making computation insufficient to hide any communication latency. \new{We quantify this using a fused GEMM implementation in \Cref{sec:fusion}.}

\subsection{Goal of this Work}
The goal of this work is to minimize \AR communication overheads during the decode phase of low-latency LLM inference using Tensor Parallelism. Unlike prior approaches that rely on compute-boundedness from large batch sizes to overlap communication with computation, our solution directly reduces \AR overheads in the low-latency inference regime, where batch sizes are minimal, and there is not enough computation to hide any communication overheads.
\section{Towards \papername}
In this section, we first describe the two \AR algorithms used in LLM inference engines. Next, we show that synchronization overheads dominate the overall runtime of \AR operations for small payloads. Finally, we present our insights that enable us to reduce the impact of these overheads by (1) eliminating the bottom-barrier via dual buffering, (2) using in-switch computation to reduce data transfer during \ar, and (3) reducing the top-barrier overhead via speculative reduction.

\subsection{\ar: Oneshot \& Twoshot Algorithms}

Most LLM inference engines, including vLLM~\cite{vllm}, SGLang~\cite{sglang}, and TRT-LLM~\cite{trtllm}, use the oneshot or the twoshot \AR algorithm~(or one of their variants) for low-latency serving. The choice depends on payload size and TP degree.  In this work we use the best of the two algorithms as our baseline.

\parhead{Oneshot.}
The oneshot \AR algorithm completes the entire reduction in a single communication round. Each GPU pulls the entire buffer from all peers and performs reduction locally. It consists of three phases: (1) a top barrier to ensure all GPUs have produced the partials before communication begins. (2) each GPU pulls the partials from all other GPUs and performs element-wise reduction (summation) on received tensors to produce the final result. (3) a bottom barrier to ensure all peers have finished reading the payload buffer before the owning GPU overwrites it.

\begin{figure}[htb!]
    \centering
    \includegraphics[width=\linewidth]{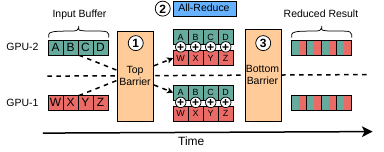}
    \caption{Oneshot \ar : each GPU pulls the entire buffer from all peers and reduces locally. Two barrier synchronizations at the start and end ensure correctness.}
    \label{fig:oneshot-allreduce}
\end{figure}

\parhead{Twoshot.}
The twoshot \AR algorithm decomposes reduction into two communication rounds: reduce-scatter followed by broadcast (all-gather). It consists of 4 phases: (1) a top barrier to ensure all GPUs have produced the partials before communication begins. (2) each GPU pulls a portion of the payload and reduces it. (3) each GPU broadcasts its reduced partial to all peers, allowing every GPU to reconstruct the full reduced result. (4) a bottom barrier to ensure that the broadcast writes are visible to all peers before moving on to the next operation. 

\begin{figure}[htb!]
    \centering
    \includegraphics[width=\linewidth]{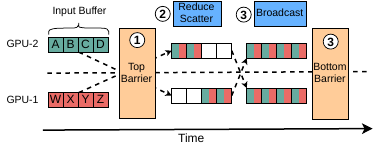}
    \caption{Twoshot \ar : each GPU reduces a portion of the payload at the switch (reduce-scatter), then broadcasts its result to all peers. Two barrier synchronizations at the start and end ensure correctness.}    
    \label{fig:twoshot-allreduce}
\end{figure}

\parhead{Data Transfer Comparison.}
The key difference between oneshot and twoshot lies in data movement. In oneshot, each GPU sends and receives $K \times (N - 1)$ bytes (where $K$ is the payload size and $N$ is the number of GPUs), scaling linearly with GPU count. In contrast, twoshot keeps data transfer constant by decomposing reduction into two steps: reduce-scatter and broadcast. The data transfer during twoshot can be further reduced using in-switch computation capabilities~\cite{ptxisa,wang2025symmetricmemory} of modern switches. With in-switch computation, across both steps of twoshot, each GPU sends $K + K/N$ and receives $K$ bytes. In this work, we focus on this in-network implementation of twoshot. Oneshot is preferred when the payload size is small and the single round trip is more beneficial. Twoshot becomes more efficient at larger scales where the reduction in data transfer outweighs the cost of the additional round trip.

\subsection{The Cost of Synchronization}
\label{sec:barrier-overhead}
\begin{figure}[htb!]
    \centering
    \includegraphics[width=\linewidth]{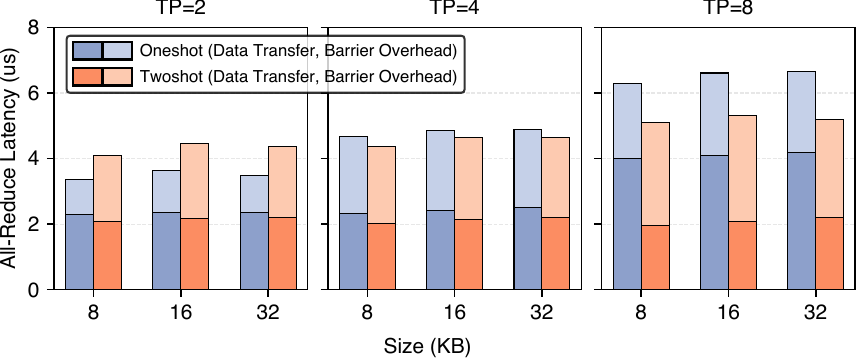}
    \caption{Breakdown of \AR latency on H200 GPUs. Synchronization overhead accounts for 32-50\% of oneshot and 49-62\% of twoshot latency for small payloads across all TP configurations.}
    \vspace{-1em}
    \label{fig:barrier-overhead}
\end{figure}

While necessary for correctness, the barriers around \AR introduce significant overhead. To quantify their impact on \AR latency, we run the oneshot and twoshot \AR implementations from vLLM~\cite{vllm-custom-ar,wang2025symmetricmemory} with and without barriers across different payload sizes and TP configurations on an H200 node. \Cref{fig:barrier-overhead} shows the results, breaking down total latency into data transfer time and synchronization time.

We observe that synchronization overhead accounts for 32-50\% of oneshot and 49-62\% of twoshot latency for small payloads (8-32~KB) across all TP configurations. For example, at TP=4 with 8~KB payload, oneshot takes 4.66~\us with barriers but only 2.33~\us without, a 2$\times$ slowdown purely from synchronization. For twoshot at TP=8 with 8~KB payload, barriers account for 3.15~\us out of 5.11~\us total, representing 62\% overhead. Twoshot exhibits higher synchronization overhead because its bottom barrier must ensure that broadcast writes are visible before the next operation, which requires a more expensive acquire-release barrier that enforces write ordering. These sizes of 4-8~KB represent the \AR payloads for most leading models~\cite{deepseekv3,deepseekr1,llama3,qwen3,gptoss,kimik2} at batch size 1. Thus, during autoregressive decoding, every \AR operation encounters this synchronization bottleneck, spending more time waiting at barriers than transferring and reducing data.

\subsection{Insight 1: Removing Bottom-Barrier via Dual Buffering}
The bottom barrier in oneshot prevents \emph{Write-After-Write (WAW)} hazards by ensuring all peers finish reading the payload buffer before the owning GPU overwrites it with new data. However, WAW is an avoidable dependency that can be eliminated through redundancy. Despite this, dual buffering is not widely used in current LLM serving systems because communication is handled by application-agnostic libraries like NCCL, which cannot make assumptions about buffer usage patterns and must include a bottom barrier to ensure correctness.

\begin{figure}[htb!]
    \centering
    \includegraphics[width=\linewidth]{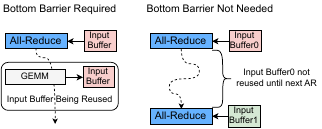}
    \caption{(left)~Application-agnostic \AR requires a bottom-barrier to prevent WAW hazards. (right)~Dual buffering ensures input buffer is reused only after the next \ar, eliminating the WAW dependency.}
    \label{fig:dual-buffering}
\end{figure}

\parhead{Eliminating the Bottom-Barrier.}
Our insight is to use \emph{dual buffering} to eliminate the bottom-barrier from \ar. By co-designing \AR with the LLM inference engine, we can safely remove this barrier. \Cref{fig:dual-buffering} illustrates our approach. We modify the inference engine to allocate two input buffers and alternate between them across successive \AR operations. This ensures that the payload buffer is never overwritten until the subsequent \AR finishes, eliminating the need for a bottom-barrier.

Dual buffering requires only 4-8~KB per token of additional HBM per GPU, which at low batch sizes is negligible compared to multi-gigabyte model weights. Despite this, it is not widely used for two reasons. First, communication libraries like NCCL~\cite{nccl} are application-agnostic and cannot make assumptions about buffer reuse patterns, so they must include a bottom barrier for correctness. Second, LLM serving systems like vLLM~\cite{vllm} employ automatic memory management that reuses tensors aggressively to minimize HBM consumption. Since the input buffer of \AR is the same size as the output of the following GEMM, the memory manager reuses it as the output buffer for the following GEMM, creating the WAW dependency shown in \Cref{fig:dual-buffering}~(left). Co-designing communication with model execution allows us to allocate a second buffer and alternate between them, breaking this dependency.

\parhead{Dual Buffering Alone is Insufficient.}
While dual buffering eliminates the bottom barrier, oneshot still scales poorly with GPU count since each GPU pulls the entire buffer from all peers. At higher TP degrees, twoshot outperforms oneshot despite requiring two round trips, because its data transfer remains constant~(see \Cref{fig:barrier-overhead}). However, the bottom barrier of twoshot enforces a Read-After-Write (RAW) dependency, ensuring broadcast writes are visible before the next operation. Unlike the WAW dependency in oneshot, this RAW dependency cannot be eliminated through dual buffering. Thus, to benefit from dual buffering, we need a oneshot design with comparable data transfer latency to twoshot.

\subsection{Insight 2: Using In-Switch Computation for Oneshot}

Modern NVIDIA switches support in-network reduction via the \texttt{ld\_reduce} instruction~\cite{ptxisa}. Each GPU maps the same virtual address range to its local buffer using symmetric memory. When a GPU issues \texttt{ld\_reduce} on a symmetric address, the switch pulls data from all GPUs, reduces it, and returns the result. Currently, \texttt{ld\_reduce} is only used for reduce-scatter, the first step of twoshot, where each GPU reduces a disjoint $K/N$ portion of the payload~(K: payload size, N: number of GPUs). However, we find that \texttt{ld\_reduce} can be used beyond reduce-scatter to lower oneshot latency.

\begin{figure}[htb!]
    \centering  
    \includegraphics[width=\linewidth]{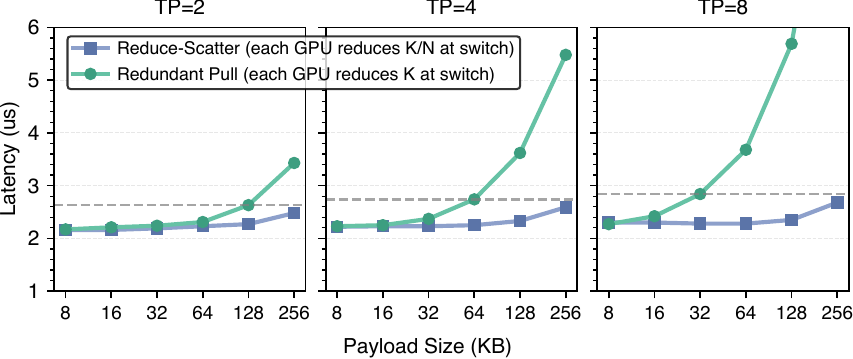}
    \caption{Latency of reduce-scatter, the first step of twoshot \ar, and redundant pull on H200 GPUs. Reduce-scatter latency remains nearly constant for small payloads, motivating in-switch reduction for the entire payload.}
    \label{fig:switch-infer}
\end{figure}

\parhead{Observation: Constant Latency for Small Payloads.}
\Cref{fig:switch-infer} shows that reduce-scatter latency remains nearly constant up to 64~KB across all TP configurations, because latency is dominated by round-trip latency between GPUs rather than data volume. At TP=8 with a 64~KB payload, reduce-scatter means each GPU reduces 8~KB at the switch, comparable to the 4-8~KB payload sizes of most leading models. Even up to 256~KB, reduce-scatter latency does not increase significantly. Our insight is that if the switch can reduce small payloads from each GPU with low latency during reduce-scatter, then each GPU can issue \texttt{ld\_reduce} over the entire payload ($K$ bytes) instead of a disjoint partition ($K/N$ bytes) and achieve data transfer latency similar to twoshot. 

\begin{figure}[htb!]
    \centering
    \includegraphics[width=0.92\linewidth]{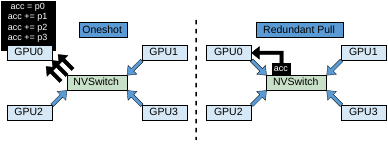}
    \caption{Redundant Pull: each GPU issues \texttt{ld\_reduce} over the entire $K$ bytes. The switch reduces data from all peers and returns the result, reducing data received per GPU from $K \times (N-1)$ to $K$.}    
    \label{fig:redundant-pull}
\end{figure}

\parhead{Redundant Pull.}
We exploit this observation to propose \emph{redundant pull}. Instead of pulling raw data from all peers and reducing locally as in oneshot, each GPU issues \texttt{ld\_reduce} over the entire buffer. The switch reduces data from all peers and returns the full result. We call this redundant pull because every GPU requests reduction on the same addresses, making the switch perform redundant reductions. \Cref{fig:redundant-pull} illustrates this design. Compared to oneshot, redundant pull reduces data received per GPU from $K \times (N-1)$ to $K$. For small payloads, redundant pull achieves data transfer latency comparable to reduce-scatter, which is only one phase of twoshot, while retaining the ability to benefit from dual buffering.

\begin{figure}[htb!]
    \centering
    \includegraphics[width=\linewidth]{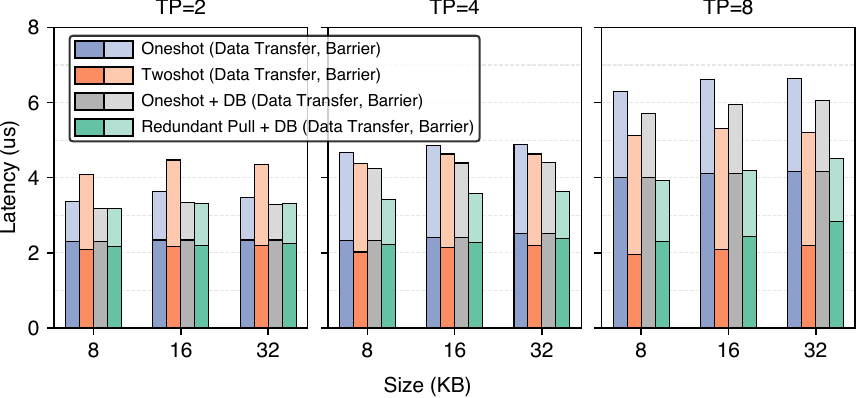}
    \caption{Latency comparison of oneshot, twoshot, oneshot with dual buffering (DB), and redundant pull with DB. Redundant pull + DB achieves the lowest latency by (1) reducing data transfer and (2) eliminating the bottom barrier.}
    \label{fig:rp-latency}
\end{figure}

\parhead{Latency Comparison.}
\Cref{fig:rp-latency} compares oneshot, twoshot, oneshot with dual buffering (DB), and redundant pull with DB. Oneshot with DB removes the bottom barrier, but its data transfer still grows with GPU count, making it slower than twoshot at TP=8. Redundant pull with DB achieves the lowest latency by combining reduced data transfer with the elimination of the bottom barrier.

\newpage
\subsection{Insight 3: Reducing Top-Barrier Overheads via Speculative Reduction}
The top barrier enforces a \emph{Read-After-Write (RAW)} dependency, ensuring all GPUs have produced valid partials before communication begins. Unlike the WAW dependency eliminated by dual buffering, this RAW dependency cannot be removed. However, we find that its overhead can be substantially reduced through \emph{speculate and verify}. To explain this, we first break down the cost of the top barrier.

\parhead{Breakdown of Barrier Costs.}
The cost of the top barrier consists of two components: \textit{divergence} and \textit{flag exchange}. Divergence is the time the fastest GPU waits for the slowest GPU to reach the barrier. Flag exchange is the time spent exchanging synchronization flags across GPUs to confirm that all participants have arrived.

\begin{figure}[htb!]
    \centering
    \includegraphics[width=\linewidth]{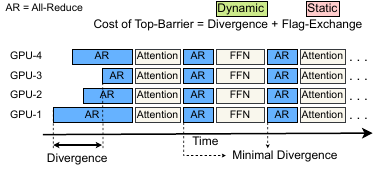}
    \caption{Representative Nsight profiler output: the cost of the top barrier decomposes into divergence (dynamic overhead) and flag exchange (static overhead). Because kernels launch via CUDA Graphs, divergence is near-zero after the first \ar, making flag exchange the dominant cost.}
    \label{fig:barrier-cost}
\end{figure}

\parhead{Divergence is Negligible in Practice.}
Modern LLM inference engines either launch all kernels for a forward pass within a single CUDA Graph~\cite{vllm,sglang}, or fuse them into a single Megakernel~\cite{spector2025megakernel}. In both cases, the forward pass executes entirely on the GPU without host interaction after the initial launch. Once GPUs are synchronized by the first \ar, they remain tightly synchronized throughout the remainder of the forward pass, since all GPUs execute identical kernels with identical input and output sizes. \Cref{fig:barrier-cost} illustrates this behavior using a representative CUDA Graph execution from Nsight profiler~\cite{nvidia_nsight_compute}, showing that divergence is near-zero after the first \ar. Most barrier time is therefore spent on flag exchange, where GPUs simply idle while exchanging synchronization flags to confirm all participants have arrived.

\parhead{Eliminating Flag-Exchange Overhead.}
Since GPUs are implicitly synchronized during the forward pass, we can \textit{speculate} that all GPUs are already synchronized and begin the data transfer phase of \AR immediately, avoiding explicit flag exchange. When a GPU enters an \ar, it speculatively begins fetching data from peers, assuming their data is ready. To ensure correctness, each GPU writes a flag to a validation buffer that is reduced alongside the payload. If the result equals $N \times \text{flag}$ after reduction, all GPUs had valid data and speculation succeeded. Otherwise, \Shortname reruns the \ar. We explain the detailed design of speculative reduction and the verification mechanism in \Cref{sec:sifar-implementation}.


\section{\shortname: Design and Implementation}
\label{sec:sifar-implementation}

\subsection{Design Overview}
\label{sec:design-overview}

\begin{figure}[htb!]
    \centering
    \includegraphics[width=\linewidth]{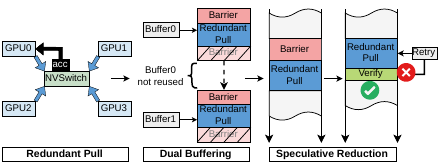}
    \caption{Overview of \shortname: (left) Redundant pull replaces oneshot by issuing \texttt{ld\_reduce} over the entire buffer to employ in-switch reduction, which reduces data transfer. (center) Dual buffering alternates two input buffers to eliminate the bottom barrier. (right) Speculative reduction removes the top barrier by speculatively beginning data transfer and verifying correctness, retrying on mis-speculation.}
    \vspace{-0.45em}
    \label{fig:spec-reduce-overview}
\end{figure}

In this section, we present the design and implementation of \emph{Synchronization Free All-Reduce (SiFAR)} which uses the optimizations described in the previous section to minimize synchronization overheads in \ar. \Cref{fig:spec-reduce-overview} illustrates our approach.

\parhead{Redundant Pull Design.}
\Shortname replaces the oneshot pull-and-reduce with redundant pull, where each GPU issues \texttt{ld\_reduce} over the entire payload buffer. The switch reduces data from all peers and returns the full result, reducing the data transfer of oneshot.

\parhead{Dual Buffering Design.}
\Shortname uses dual buffering to remove the bottom barrier by maintaining two buffers for consecutive \AR operations. Thus, the input buffer to an \AR is not reused until after the next \ar, where the GPUs are again synchronized, safely eliminating the bottom barrier. We explain how \Shortname integrates dual buffering with the automatic memory management of vLLM in \Cref{sec:implementation-details}.

\parhead{Speculative Reduction Design.}
\Shortname minimizes the overhead of the top barrier using speculative reduction. When a GPU enters the \AR kernel, it speculatively begins issuing
\texttt{ld\_reduce} to fetch reduced data from remote peers, assuming all peer GPUs have already finished the previous operation. To facilitate validation, \Shortname allocates a small validation buffer per threadblock.
After the preceding operation writes its result to the payload buffer, each GPU writes a flag to the validation buffer. The flag is reduced alongside the payload. If the reduced validation result equals \texttt{ngpus $\times$ flag}, all GPUs had valid data and speculation succeeded. Otherwise, \Shortname reruns the \AR until validation succeeds.

\begin{figure}[htb!]
    \centering
    \includegraphics[width=0.85\linewidth]{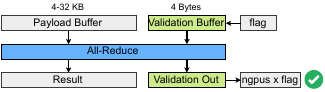}
    \caption{Verification Logic to ensure correctness of speculative reduction. Each GPU writes a flag to a validation buffer before beginning the \ar. After reduction, the reduced validation result is checked to detect mis-speculation.}
    \vspace{-2em}
    \label{fig:verification-logic}
\end{figure}

\subsection{Implementation Details}
\label{sec:implementation-details}

\parhead{Redundant Pull.}
We replace oneshot pull-and-reduce with redundant pull by changing how each threadblock performs reduction:

\begin{lstlisting}[style=pythonstyle]
# Oneshot: pull from each peer and reduce locally
for peer in range(ngpus):
    data = load(peer, payload_buf)
    result += data
# Redundant Pull: switch reduces and returns result
result = mutlimem.ld_reduce(payload_buf)
\end{lstlisting}

\parhead{Dual Buffering.}
Implementing dual buffering requires controlling the allocation and reuse of buffers. However, frameworks like vLLM and SGLang rely on automatic memory management in Pytorch, which eagerly reclaims buffers. Since the input buffer to \AR matches the output size of the next GEMM, it is quickly reused:
\begin{lstlisting}[style=pythonstyle]
buf = redundant_pull(input_buf0)
input_buf0 = GEMM(buf, weight)  # Buffer reused
...
buf = redundant_pull(input_buf1)
\end{lstlisting}

To implement dual buffering without modifying framework internals, we modify the \AR function signature to accept an additional argument referencing the previous input buffer:

\begin{lstlisting}[style=pythonstyle]
buf = redundant_pull(input_buf0)
...
buf = redundant_pull(input_buf1, input_buf0)
\end{lstlisting}

By passing \texttt{input\_buf0} as an additional argument to the second \ar, the memory manager is forced to keep \texttt{input\_buf0} alive until after the second \AR executes, since it cannot guarantee that \texttt{input\_buf0} will be unused. The additional argument is retained solely for its memory management side-effect and is not used in the computation itself. This approach requires only minimal changes to model execution code and works across frameworks without modifying their memory management systems.

\parhead{Speculative Reduction.}
To implement speculative reduction, we build the \Shortname \AR kernel using redundant pull for the payload and oneshot pull-and-reduce for validation. We show a pseudocode implementation below:

\begin{lstlisting}[style=pythonstyle]
int* valid_buf;  # Dual buffered
valid_buf[blockIdx.x] = flag;
__syncthreads();
result = switch_reduce(payload_buf);  # warps 0..k
valid_out = local_reduce(valid_buf);  # warp k+1
while (valid_out != ngpus x flag) {
  result = switch_reduce(payload_buf);
  # Use .cg to bypass L1 cache for valid_buf
  valid_out = local_reduce_cg(valid_buf);
    __syncthreads();
}
flag = flag + 1
\end{lstlisting}

The payload buffer is reduced using \texttt{switch\_reduce}, which performs reduction in the switch. For the validation buffer, we instead use pull-and-local-reduce to compute \texttt{valid\_out}, as it empirically outperforms using \texttt{switch\_reduce} for this purpose. Each threadblock runs with one extra warp to process the validation buffer while the remaining warps handle the payload. Before beginning the \ar, each GPU writes the \texttt{flag} value into its validation buffer. After reduction, we check if the result matches \texttt{ngpus $\times$ flag}. If validation fails, we re-execute until validation succeeds. 

\parhead{Verification Logic Correctness.}
To maintain correctness across multiple \AR invocations, we increment the flag value after each \ar, ensuring each \AR has a unique validation result. We apply dual buffering to the validation buffer to prevent race conditions where a fast GPU might overwrite its flag before all peers have finished reading it. Since remote loads are cached in L1~\cite{fusco2024understanding}, upon mis-speculation we use the \texttt{.cg} modifier to load the validation buffer to bypass L1 and fetch fresh data. The payload \texttt{switch\_reduce} does not require this, since data is fetched from the switch and not cached locally. The correctness of the verification logic relies on GPU monotonicity; see \Cref{sec:correctness-safety} for details.


\parhead{Megakernel Integration.}
In Megakernels~\cite{spector2025megakernel}, the entire forward pass is fused into a single kernel, and successive operations within the kernel communicate completion via flags. \Shortname reuses these existing operator flags for validation by reading and reducing the flag from all remote GPUs, instead of explicitly writing to a separate validation buffer. This eliminates the overhead of writing to the validation buffer before each \ar, further reducing latency.

\section{Evaluation Methodology}
\label{sec:eval-methodology}

\parhead{Serving Framework.}
We evaluate \Shortname by integrating it into the Megakernel framework~\cite{spector2025megakernel}. Megakernels fuse the entire forward pass into a single CUDA kernel, eliminating kernel launch overheads and improving HBM utilization. Within the Megakernel, weights of the next operation are prefetched into shared memory whenever possible, including during \ar, further reducing latency. We evaluate on Megakernels rather than serving systems like vLLM~\cite{vllm} because Megakernels represent the state-of-the-art for low-latency inference, where non-communication overheads are already minimized, making the impact of \AR optimization most visible. We extend the original Megakernel implementation to support Llama-3.1-8B and Qwen3.5-397B-17B with multi-GPU tensor parallelism. Our Megakernel baseline without \Shortname already achieves 1.3$\times$ lower TPOT than vLLM~0.18.1 for Llama-3.1-8B and 2.2$\times$ for Qwen3.5-397B-17B at TP=8, confirming that non-communication overheads are substantially reduced. \new{We run both models on batch size 1 unless otherwise specified.}

\parhead{Models.}
We evaluate \Shortname on two models: Llama-3.1-8B~\cite{llama3}, a dense transformer, and Qwen3.5-397B-17B~\cite{qwen3.5}, a Mixture-of-Experts (MoE) model that activates only 17B parameters per token but requires storing all 397B parameters in memory. We run both models in FP8 precision. Due to model size, Qwen3.5-397B-17B is evaluated only at TP=4 and TP=8. We evaluate \Shortname using input context lengths from 1K to 16K tokens.

\parhead{Hardware.}
We use a single node with 8$\times$ NVIDIA H200 GPUs, each with 141~GB of HBM3e memory. We use CUDA 12.9 with default clock frequencies. We evaluate TP across 2, 4, and 8 GPUs.

\parhead{Metrics.}
We measure standalone \AR latency to isolate the impact of \shortname. For end-to-end impact, we measure throughput (inverse of TPOT) in tokens per second during the decode phase. We report speedup relative to the baseline implementation.

\parhead{Baselines.}
Our baseline uses the best of oneshot and twoshot \AR for each configuration. We also compare against two variants of oneshot: TRT-LLM oneshot~\cite{trtllm} and Lamport oneshot~\cite{chandy1985distributed}. We describe these in detail in \Cref{sec:evaluation}.

\begin{figure*}[htb!]
    \centering
    \includegraphics[width=\linewidth]{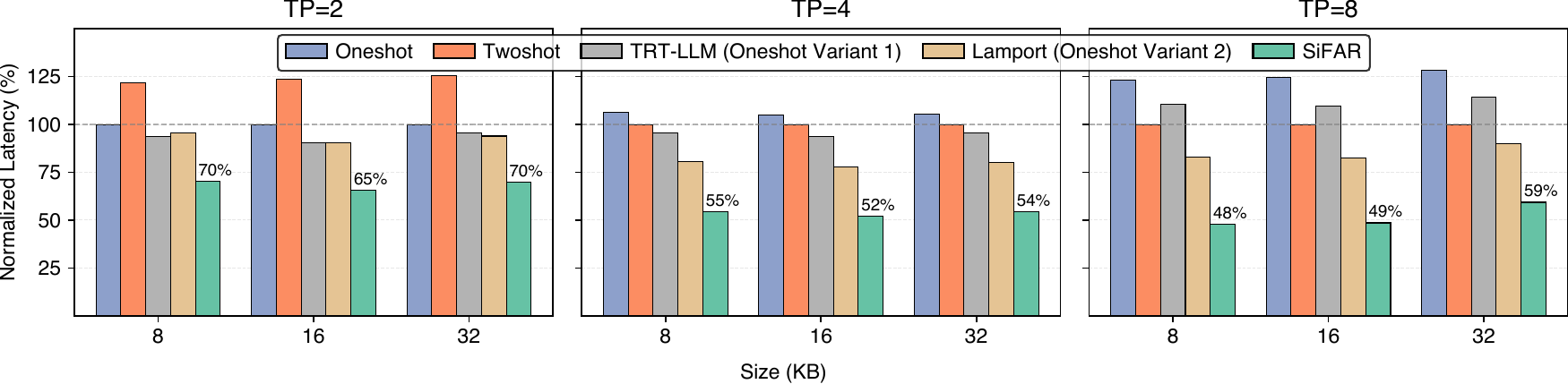}
    \caption{\AR latency across five implementations for TP=2, 4, and 8 with payloads from 8 to 32~KB, normalized to the best of oneshot and twoshot (100\%). \Shortname consistently outperforms all other implementations, achieving up to 52\% latency reduction at TP=8 by combining redundant pull, dual buffering, and speculative reduction.}
    \label{fig:latency-comparison}
\end{figure*}

\newpage
\section{Evaluation}
\label{sec:evaluation}
In this section, we evaluate \shortname. We first analyze its impact on \AR latency, then measure end-to-end LLM inference improvements, and finally study the contribution of each optimization.

\subsection{Impact on \AR Latency}

\Cref{fig:latency-comparison} compares standalone \AR latency across five implementations: oneshot, twoshot, TRT-LLM, Lamport, and \Shortname for TP configurations of 2, 4, and 8 GPUs with payloads from 8 to 32~KB. The baseline for each configuration is the best of oneshot and twoshot, which is the default in vLLM and SGLang. We execute \AR back-to-back within a Megakernel to avoid measuring kernel launch overheads. The measured latency includes any mis-speculation overheads incurred by \shortname.

\Shortname consistently outperforms all other implementations across every configuration. At TP=2, where oneshot is the best baseline, \Shortname reduces latency to 2.36~\us compared to 3.36~\us for oneshot at 8~KB, a 30\% reduction. At TP=4, twoshot becomes the best baseline and \Shortname achieves a 45\% reduction in latency from 4.38~\us to 2.39~$\mu$s. Finally, the benefits of \Shortname are largest at TP=8, where the baseline is twoshot with 5.11~\us latency at 8~KB, and \Shortname reduces it to 2.44~\us, a 2$\times$ improvement. Next, we compare \Shortname against two variants of oneshot: TRT-LLM oneshot and Lamport oneshot.

\parhead{TRT-LLM (Oneshot).}
We compare \Shortname against the TRT-LLM oneshot implementation, which operates in 3 phases: (1)~each GPU copies its input data into a separate local buffer, making the payload buffer private to the \AR kernel so the bottom barrier can be safely removed. (2)~A top barrier ensures all GPUs have produced valid partials. (3)~Each GPU pulls the full buffer from all peers and performs local reduction. Since TRT-LLM still pulls and reduces like oneshot, receiving $K \times (N-1)$ bytes per GPU, it does not scale well to higher TP degrees. \Shortname outperforms TRT-LLM by 25-28\% at TP=2, 43-44\% at TP=4, and 48-57\% at TP=8.

\parhead{Lamport (with In-Network Multicast)~\cite{chandy1985distributed}.}
Finally, we compare \Shortname against an optimized implementation (using In-Network Multicast) of a Lamport-style \ar, which is a single-round, push-based algorithm. It operates in 3 phases: (1)~each GPU pushes its partial results to all peers via \texttt{multimem.st}, which multicasts data through NVSwitch so each GPU sends only $K$ bytes. (2)~A bottom barrier, similar to the one in twoshot, ensures that all multicast writes are visible locally. (3)~Each GPU then loads all received partials from its local buffer and reduces them. 
Lamport avoids the top barrier and transfers the same total data as redundant pull, but in the opposite direction: it sends $K$ bytes and receives $K \times N$ bytes into local memory. In contrast, redundant pull sends $K \times N$ bytes and receives only $K$ bytes of reduced data directly into registers via \texttt{multimem.ld\_reduce}. \Shortname outperforms Lamport by 32-33\% at TP=4 and 34-43\% at TP=8 by avoiding the acquire-release bottom barrier and the on-chip latency of loading the full payload from local memory before reduction.

\begin{figure}[htb!]
    \vspace{-0.2em}
    \centering
    \includegraphics[width=\linewidth]{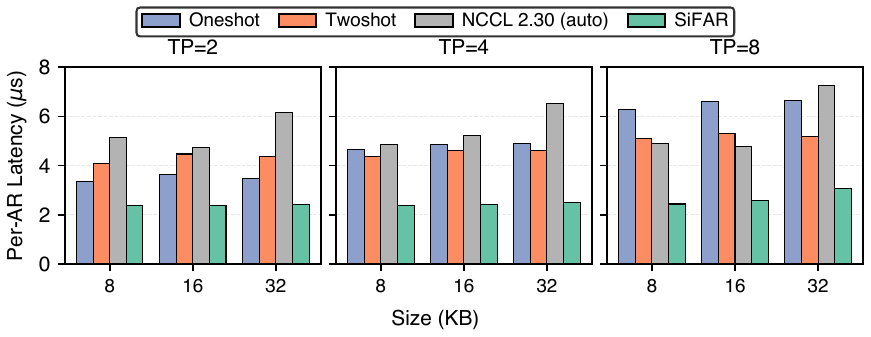}
    \caption{\new{\AR latency with NCCL 2.30 auto-selection for TP=2, 4, and 8 across 8--32~KB payloads. NCCL is competitive with the best of oneshot and twoshot, but \Shortname achieves the lowest latency across all configurations.}}
    \label{fig:nccl-comparison}
\end{figure}

\parhead{Comparison with NCCL 2.30.}
\new{
We compare \Shortname against NCCL 2.30, which provides custom low-latency kernels that use the LL protocol~\cite{hu2025demystifying}. We run NCCL with its default auto-selection policy, which uses a heuristic to choose a communication strategy based on the network topology and payload size. \Cref{fig:nccl-comparison} shows that NCCL auto is only better than the best of oneshot and twoshot for the 8 and 16~KB payloads at TP=8. Finally, despite the custom low-latency kernels, \Shortname achieves 46--62\% lower latency than NCCL across evaluated TP degrees and payload sizes.}

\subsection{Impact on End-to-End LLM Inference}
\begin{figure}[htb!]
    \centering
    \includegraphics[width=\linewidth]{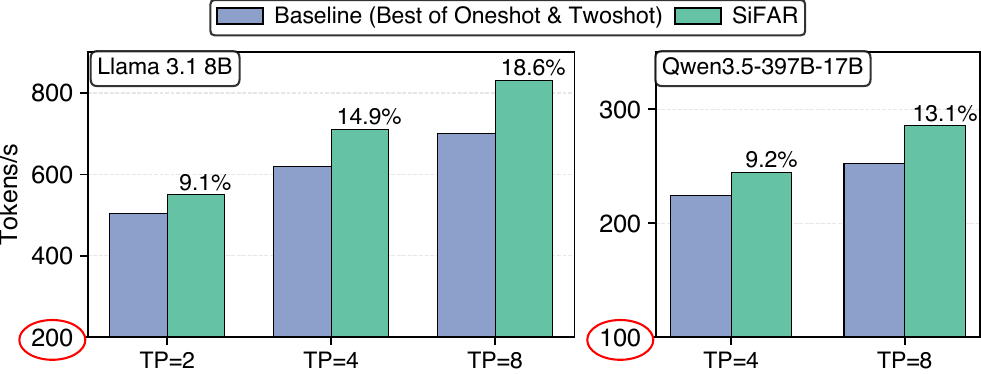}
    \caption{End-to-end throughput (tokens/s) for Llama-3.1-8B and Qwen3.5-397B-17B across TP configurations. \Shortname improves throughput by \replace{7.0\%}{9.1\%} to \replace{15.2\%}{18.6\%} for Llama-3.1-8B and up to \replace{12.2\%}{13.1\%} for Qwen3.5-397B-17B.}
    \label{fig:e2e-results}
\end{figure}

\Cref{fig:e2e-results} shows end-to-end throughput across TP degrees for Llama-3.1-8B and Qwen3.5-397B-17B, with input and output sequence length of 1000. \Shortname improves throughput consistently across all configurations \new{with best of oneshot and twoshot as the baseline}. For Llama-3.1-8B, the improvement grows with TP degree: \replace{7.0\%}{9.1\%} at TP=2, \replace{12.4\%}{14.9\%} at TP=4, and \replace{15.2\%}{18.6\%} at TP=8. For Qwen3.5, we observe a similar trend with \replace{8.7\%}{9.2\%} improvement at TP=4 and \replace{12.2\%}{13.1\%} at TP=8. This scaling behavior is expected because the impact of \AR increases with TP degree, making the optimizations of \Shortname more impactful.

\begin{figure}[htb!]
    \centering
    \includegraphics[width=\linewidth]{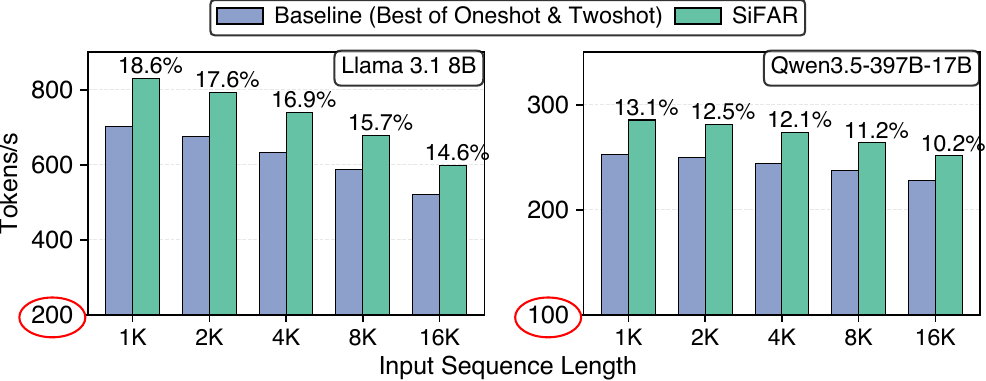}
    \caption{\new{End-to-end throughput (tokens/s) for Llama-3.1-8B and Qwen3.5-397B-17B at TP=8 across input sequence lengths and output sequence length = 1000. Improvement decreases with longer sequences as attention dominates, but \Shortname still provides 14.6\% and 10.2\% improvement at ISL=16K.}}
    \label{fig:e2e-isl-ablation}
\end{figure}

\parhead{Impact of Input Sequence Length (ISL).}
\Cref{fig:e2e-isl-ablation} shows the impact of \Shortname when varying ISL for Llama-3.1-8B and Qwen3.5-397B-17B at TP=8. For Llama-3.1, \Shortname provides a 18.6\% improvement at ISL=1000 and 14.6\% at ISL=16K. For Qwen3.5, the improvement ranges from 13.1\% at ISL=1000 to 10.2\% at ISL=16K. As ISL increases, the fraction of time spent in attention grows, which does not benefit from \shortname. However, even at ISL=16K, \Shortname provides meaningful improvements because \AR remains a significant fraction of total execution time.

\begin{figure}[htb!]
\centering
\includegraphics[width=\linewidth]{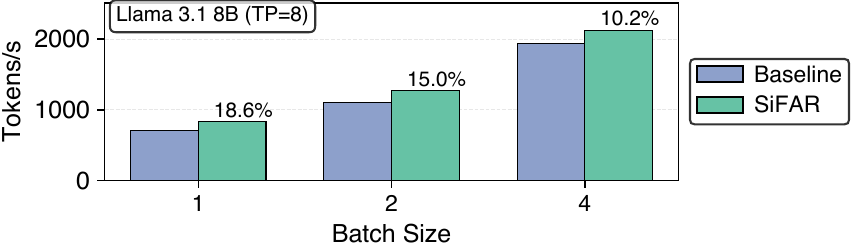}
\caption{\new{Batch-size sensitivity for Llama-3.1-8B at TP=8. \Shortname improves throughput by 18.6\% at BS=1, and the gain shrinks to 10.2\% at BS=4 as larger batches make \AR a smaller fraction of TPOT.}}
\vspace{-1em}
\label{fig:bs-eval}
\end{figure}

\parhead{Impact of Batch Size.}
\new{
\Cref{fig:bs-eval} evaluates \Shortname across BS=1, 2, and 4 on Llama-3.1-8B at TP=8. \Shortname improves throughput by 18.6\% at BS=1, 15.0\% at BS=2, and 10.2\% at BS=4. The gain is largest at BS=1 and shrinks as the batch grows. Larger batches increase the GEMM and attention compute, so \AR is a smaller fraction of TPOT. \Shortname targets the BS=1 regime because it is latency-critical. Although larger batches raise throughput, they worsen per-request TPOT by 45\% from BS=1 to BS=4.
}

\begin{figure}[htb!]
\centering
\includegraphics[width=\linewidth]{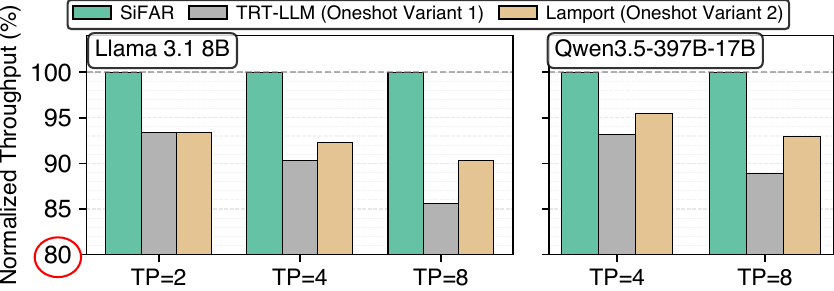}
\caption{\new{End-to-end throughput comparison across \AR implementations, normalized to \Shortname. \Shortname outperforms the TRT-LLM oneshot variant and the Lamport-style push-based \ar.}}
\label{fig:e2e-alternative}
\end{figure}

\parhead{Comparison with Other \AR Algorithms.}
\new{
\Cref{fig:e2e-alternative} compares the throughput of \Shortname against the TRT-LLM oneshot variant and the Lamport-style push-based. At TP=8 for Llama-3.1-8B, TRT-LLM and Lamport reach only 85.6\% and 90.3\% of \Shortname throughput, so \Shortname outperforms even these optimized oneshot-style implementations end-to-end.
}

\begin{figure}[htb!]
\centering
\includegraphics[width=\linewidth]{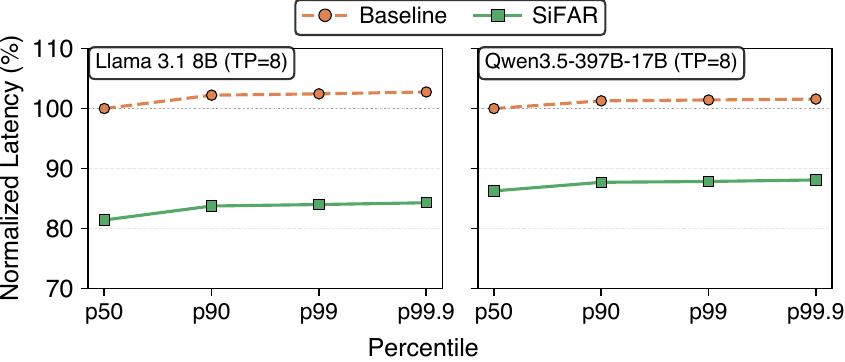}
\caption{\new{End-to-end TPOT percentiles for Llama-3.1 and Qwen3.5 at TP=8, normalized to the baseline p50 latency. The speedup from \Shortname holds through the tail, reducing p99 latency to 84.0\% for Llama-3.1-8B and 87.8\% for Qwen3.5-397B-17B, so mis-speculation retries do not inflate tail latency.}}
\label{fig:e2e-p99}
\end{figure}

\parhead{Impact of \Shortname on Tail Latency.}
\new{
\Cref{fig:e2e-p99} reports TPOT percentiles at TP=8, normalized to the baseline p50. \Shortname lowers both median and tail latency: for Llama-3.1-8B it reduces p50 to 81.4\% and p99 to 84.0\%, and for Qwen3.5-397B-17B it reduces p50 to 86.3\% and p99 to 87.8\%. The speedup from \Shortname holds through p99.9, so mis-speculation retries do not inflate tail latency.
}

\subsection{Impact of Individual Optimizations}
\begin{figure}[htb!]
    \centering
    \includegraphics[width=0.9\linewidth]{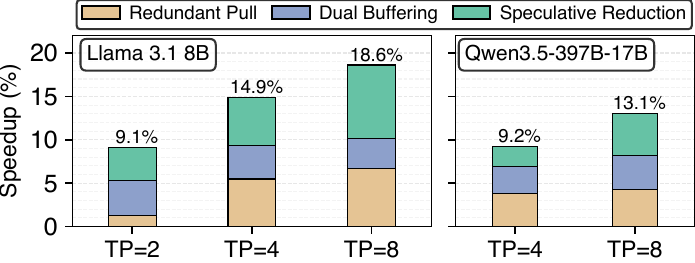}
    \caption{Breakdown of the speedup from our 3 optimizations for Llama-3.1-8B and Qwen3.5-397B-17B across TP configurations. All three optimizations contribute meaningfully, with benefits increasing at higher TP degrees.}
    \label{fig:e2e-speedup-breakdown}
\end{figure}

To understand the individual contributions of our optimizations, we conduct an ablation study where we enable each optimization incrementally: redundant pull alone, redundant pull with dual buffering, and \Shortname with all three optimizations including speculative reduction. \Cref{fig:e2e-speedup-breakdown} shows the contributions of each of our optimizations.

For Llama-3.1-8B, redundant pull contributes \replace{5.2\%}{6.7\%} at TP=8 by reducing data transfer through in-switch reduction. Dual buffering builds on this by eliminating the bottom barrier, contributing an additional \replace{5.6\%}{3.4\%} at TP=8. Speculative reduction reduces latency by minimizing the top-barrier overhead, contributing \replace{3.0\%}{3.8\%} at TP=2 and up to \replace{4.4\%}{8.5\%} at TP=8. Together, the three optimizations are complementary and address distinct sources of \AR overhead.

\begin{figure}[htb!]
    \centering
    \includegraphics[width=\linewidth]{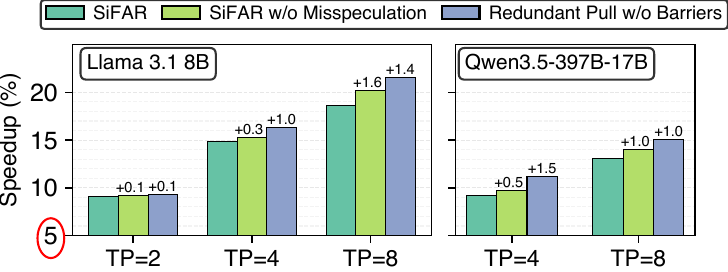}
    \caption{\Shortname compared against idealized configurations: perfect speculation (no mis-speculation) and redundant pull with no barriers. The gaps quantify the cost of mis-speculation retries and verification overhead.}
    \label{fig:e2e-speedup-comparison}
\end{figure}

\parhead{Mis-Speculation and Verification Overhead.}
We compare \Shortname against two idealized configurations in \Cref{fig:e2e-speedup-comparison}: perfect speculation (no retries) and redundant pull without barriers. The gap between \Shortname and perfect speculation isolates the cost of mis-speculation retries, while the gap between perfect speculation and redundant pull without barriers isolates the verification overhead.

\begin{figure}[htb!]
    \centering
    \includegraphics[width=0.9\linewidth]{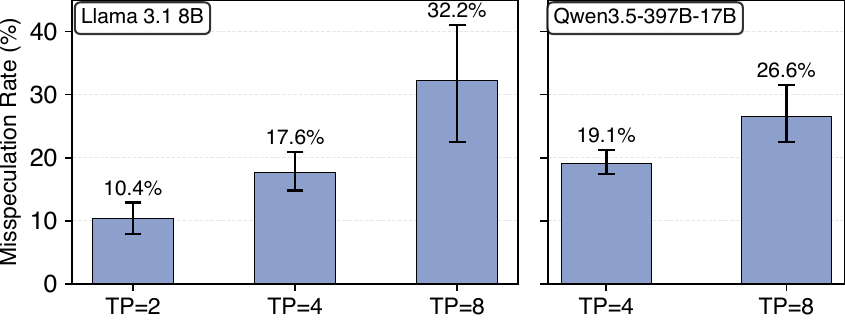}
    \caption{Mis-speculation rate for Llama-3.1-8B and Qwen3.5-397B-17B. While mis-speculation rate increases with TP degree, it results in at most \replace{1.4\%}{1.6\%} lost performance.}
    \label{fig:e2e-misspec-rate}
    \vspace{-1em}
\end{figure}

For Llama-3.1-8B at TP=8, \Shortname achieves a \replace{15.2\%}{18.6\%} improvement. With perfect speculation, the improvement increases to \replace{16.3\%}{20.2\%}, meaning mis-speculation retries cost only \replace{1.1\%}{1.6\%} despite a \replace{34.2\%}{32.2\%} average mis-speculation rate~(\Cref{fig:e2e-misspec-rate}). Removing barriers entirely yields \replace{17.6\%}{21.6\%}, so verification overhead accounts for an additional \replace{1.3\%}{1.4\%}. Together, mis-speculation and verification account for only \replace{2.4\%}{3.0\%} of lost improvement, showing that speculative reduction captures most of the available benefit.

\begin{figure}[htb!]
    \centering
    \includegraphics[width=\linewidth]{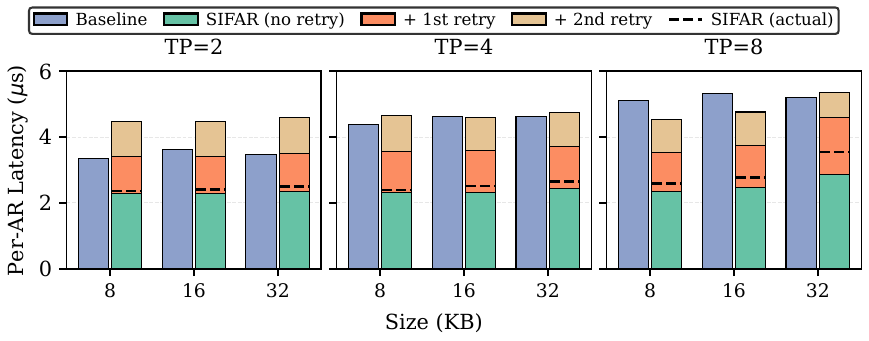}
    \caption{\new{Retry latencies of \shortname. Even with retries, the average \Shortname latency stays below the baseline.}}
    \label{fig:sifar-retry}
\end{figure}

\new{
\parhead{Retry Overhead.}
\Cref{fig:sifar-retry} shows the \Shortname latency for the no-retry case and the added latency from each retry. We measure this by forcing \Shortname to retry a fixed number of times. Even with retries, the average \Shortname latency stays well below the baseline (oneshot or twoshot) across all evaluated TP degrees and payload sizes. Each retry costs less than the first attempt, which pays the one-time startup cost of beginning the \ar.
}

\subsection{Comparison with Compute-Comm. Fusion}
\label{sec:fusion}
\begin{figure}[htb!]
\centering
\includegraphics[width=\linewidth]{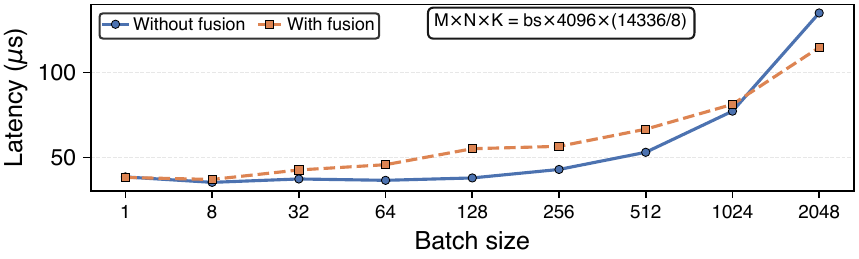}
\caption{\new{Compute-communication fusion from ParallelKittens for the down-projection GEMM of Llama-3.1-8B at TP=8. Fusion helps only at large batch sizes, while small batches have insufficient computation to hide communication.}}
\label{fig:fusion}
\vspace{-1em}
\end{figure}

\new{
\Cref{fig:fusion} compares fine-grained compute-communication fusion from ParallelKittens~\cite{sul2025parallelkittens} against an unfused implementation for the down-projection GEMM of Llama-3.1-8B at TP=8. At small batch sizes, the fused GEMM has similar or worse latency, because there is not enough computation to overlap with communication. Fusion helps only at very large batch sizes, such as BS=2048, where the GEMM has enough work to hide communication. In the regime that \Shortname targets, overlap provides little benefit, so \Shortname reduces \AR latency directly instead of trying to hide it.
}

\clearpage
\section{Related Work}

We discuss related work in four categories: (1) speculation and relaxation techniques to reduce synchronization overheads, (2) low-latency LLM inference optimizations, (3) collective communication optimizations, and (4) in-network push-based reduction.

\parhead{Using Speculation to Reduce Synchronization Overheads.}
A long line of work explores reducing synchronization latency by executing critical sections speculatively rather than waiting at locks or barriers. Early hardware proposals such as Transactional Memory (TM)~\cite{herlihy1993transactional}, Speculative Lock Elision~\cite{rajwar2001speculative}, and Speculative Synchronization~\cite{martinez2002speculative} execute code regions in parallel while deferring conflict checks to commit time. TM provides architectural support for atomic sections whose updates are speculatively buffered and committed only when no conflicts are detected. Transactional Lock-Free Execution~\cite{rajwar2002transactional} extends this model using timestamp-based ordering to allow multiple threads to speculatively enter the same critical section. Subsequent work, such as TMbarrier~\cite{pedrero2018tmbarrier} and Elastic Barriers~\cite{tiwari2025unleashing}, applies similar ideas to barrier synchronization by speculating across phase boundaries. Speculative Lock Forwarding~\cite{yaghini2019speclock} and Redundant Barrier Elimination~\cite{chabbi2015barrier} further reduce waiting by dynamically validating safety conditions at runtime.

While these techniques also use speculation to reduce synchronization costs, they primarily target removing locks and barriers placed conservatively for cases where threads may not actually conflict. In contrast, \Shortname addresses scenarios with true read-after-write dependencies where multiple GPUs must be synchronized for correctness. \Shortname exploits the observation that GPUs executing identical work remain largely synchronized, allowing it to speculate past the barrier and eliminate explicit synchronization latency.

\parhead{Low-Latency LLM Inference Optimizations.}
Programmatic Dependent Launch (PDL)~\cite{pdl_docs} reduces inter-kernel idle time by allowing a kernel to begin while the preceding one is still running. Megakernels~\cite{spector2025megakernel} fuse the entire forward pass into a single kernel, eliminating launch overheads and enabling weight prefetching. Mirage~\cite{wu2024mirage} provides a compiler that automatically fuses operators into Megakernels, making them easier to implement. Speculative Decoding~\cite{speculative-decoding} reduces the number of sequential decode steps by generating multiple tokens in parallel. \Shortname is complementary to all of these: it reduces the latency of \AR communication within each decode step, which these techniques do not address.

\parhead{Collective Communication Optimizations.}
During training, \AR communication accounts for a major fraction of training time due to the large volume of gradient data exchanged. Multiple works exploit tolerance to gradient errors by communicating only a subset of gradients~\cite{kasan2025skipreduce, li2022near, fei2021efficient}, applying lossy compression~\cite{vogels2019powersgd, li2024thc}, or using quantization~\cite{renggli2019sparcml, alistarh2017qsgd, aji2017sparse}. Our work is orthogonal as we focus on low-latency inference, where the payload size is very small, and the latency of the synchronization barrier is significant. During inference, PRESERVE~\cite{yuzuguler2025preserve} proposes prefetching model weights from HBM to L2 cache during \AR operations to overlap memory access with communication. However, our evaluation setup using Megakernels already prefetches weights into shared memory, and for MoE models, prefetching during \AR is not possible because expert selection depends on the \AR output. \Shortname directly reduces \AR latency rather than hiding it, providing benefits even when prefetching opportunities are exhausted.

\parhead{In-Network Push-Based Reduction.}
Several works propose push-based in-network reduction where GPUs send data to a switch, which reduces and broadcasts without a top barrier. SwitchML~\cite{sapio2021scaling} uses programmable switches for streaming aggregation during training. Klenk et al.~\cite{klenk2020network} propose push-based reduction in NVSwitch fabrics, achieving up to 18$\times$ speedup for small messages. CAIS~\cite{zhang2026towards} extends NVSwitch with compute-aware request merging that unifies push and pull. These approaches require hardware modifications to the switch. In contrast, \Shortname achieves barrier-free reduction on current systems by combining dual buffering and speculative reduction, without requiring push-based hardware support.

\new{
\section{Discussion}

\subsection{Monotonicity Assumption and Correctness}
\label{sec:correctness-safety}
\shortname's speculative reduction relies on a validation mechanism to ensure correctness, which depends on the \emph{monotonicity assumption}. Consider the case where GPU1 is ahead of GPU0, so GPU1 reads GPU0's payload before it finishes writing. In this case, the verification flag correctly indicates an error. However, if GPU1 slows down after reading the payload and only reads the verification flag after GPU0 has caught up, GPU1 would read an incorrect payload but a correct flag, causing incorrect verification. Thus, correctness assumes that if GPU1 is ahead of GPU0 at time $T_1$, it remains ahead at time $T_2 > T_1$, i.e., relative GPU progress is monotonic.

In LLM inference, this assumption holds for two reasons. First, CUDA Graphs or Megakernels eliminate host-side variability, the primary source of extreme GPU divergence. Second, GPU kernels typically execute with a run-to-completion model. Unlike CPU processes that can be easily preempted, GPU kernels run atomically to completion unless explicitly preempted~\cite{nvidia-pascal-preemption,davies2025kitsune}. Thus, once a kernel begins executing, its relative progress with respect to other kernels is unlikely to change significantly, i.e., if GPU1 is ahead of GPU0 when reading the payload, it will likely remain ahead when reading the verification flag. To verify this independently of the validation mechanism itself, we compare the output of \Shortname against the twoshot \ar. Across 100,000 decode iterations spanning multiple models and TP configurations, \Shortname matches this baseline, confirming that no silent error occurs.

}



\section{Conclusion}
With the rise of reasoning models and agentic systems, token generation latency has become critical for LLMs. To minimize latency, inference engines reduce batching and use Tensor Parallelism to leverage aggregate memory bandwidth across GPUs. However, \AR emerges as the dominant cost as other bottlenecks diminish. Removing \AR improves throughput by \replace{35\%}{43\%} at TP=8 for Llama~3.1~8B on H200 GPUs. We present \fullname, which minimizes \AR overhead through three techniques: redundant pull to reduce data transfer via in-switch computation, dual buffering to eliminate the bottom barrier, and speculative reduction to minimize the top barrier. We integrate \Shortname into Megakernels, a state-of-the-art approach for low-latency inference, and evaluate it on Llama-3.1-8B and Qwen3.5-397B-17B. \Shortname reduces \AR latency by up to 52\% and improves end-to-end throughput by \replace{15.2\%}{18.6\%} for Llama-3.1-8B and \replace{12.2\%}{13.1\%} for Qwen3.5-397B-17B at TP=8.



\begin{acks}
We thank the anonymous reviewers of ISCA-2026 and MICRO-2026 for their valuable feedback and suggestions. This work was supported, in part, by NSF grant 233304.
\end{acks}


\bibliographystyle{ACM-Reference-Format}
\bibliography{refs}

\end{document}